\documentclass[preprint,prd,aps,tighten,nofootinbib,amssymb]{revtex4}
\usepackage{color}
\input{colordvi.tex}
\usepackage{amsmath}
\usepackage[dvips]{graphicx}
\usepackage{subfigure}
\usepackage{comment}
\newcommand{\beq}{\begin{equation}}
\newcommand{\eeq}{\end{equation}}
\newcommand{\beqa}{\begin{eqnarray}}
\newcommand{\eeqa}{\end{eqnarray}}

\newcommand{\bea}{\begin{eqnarray}}
\newcommand{\eea}{\end{eqnarray}}
\newcommand{\bear}{\begin{array}}
\newcommand {\eear}{\end{array}}
\newcommand{\bef}{\begin{figure}}
\newcommand {\eef}{\end{figure}}
\newcommand{\bec}{\begin{center}}
\newcommand {\eec}{\end{center}}
\newcommand{\non}{\nonumber}

\def\REF#1{(\ref{#1})}

\def\lrf#1#2{ \left(\frac{#1}{#2}\right)}
\def\lrfp#1#2#3{ \left(\frac{#1}{#2} \right)^{#3}}

\begin{document}
\widetext
\draft

\begin{flushright}
RIKEN-MP-53\\
TU-917
\end{flushright}

\title{Dark Radiation and Dark Matter \\
in Large Volume Compactifications
}

\author{Tetsutaro Higaki}%
\affiliation{Mathematical Physics Lab., RIKEN Nishina Center, Saitama 351-0198, Japan}
\author{Fuminobu Takahashi}
\affiliation{Department of Physics, Tohoku University, Sendai 980-8578, Japan}

\date{\today}

\begin{abstract}
We argue that dark radiation is naturally generated from the decay of the overall volume modulus 
in the LARGE volume scenario. We consider both sequestered and non-sequestered cases, and
find that the axionic superpartner of the modulus is produced by the modulus decay and it 
can account for the dark radiation suggested by observations, while the
modulus decay through the Giudice-Masiero term
gives the dominant contribution to the total decay rate.
In the sequestered case, the lightest supersymmetric particles produced by the modulus decay 
can naturally account for the observed dark matter density. 
In the non-sequestered case, on the other hand, 
the supersymmetric particles are not produced by the modulus decay, since
the soft masses are of order the heavy gravitino mass. The QCD axion will then be a plausible dark matter candidate.
\end{abstract}

\maketitle
\section{Introduction}
\label{sec:1}

In superstring theories, moduli fields necessarily appear at low energies through 
compactifications. 
Supersymmetric compactifications, e.g., on a Calabi-Yau (CY) space \cite{Candelas:1985en},
naturally lead to massless moduli and their axionic superpartners 
at the perturbative level because 
of the remnant of higher dimensional gauge symmetry: 
\begin{align}
T_{\rm moduli} \;\rightarrow\;  T_{\rm moduli}  + i \alpha,
\label{shiftsym}
\end{align}
where $\alpha$ is a real transformation parameter. 

Most of these moduli must be stabilized  in order to get a sensible low-energy theory, 
since the moduli determine all the physically relevant quantities such as the
size of the extra dimensions, physical coupling constants, and even the supersymmetry (SUSY) breaking
scale. To this end, closed string flux backgrounds in extra dimensions, 
i.e. flux compactifications \cite{Grana:2005jc, Blumenhagen:2006ci}, 
are powerful tools to fix many moduli simultaneously. Most of the remaining moduli which are not stabilized by 
the fluxes can become massive by instantons/gaugino condensations like in the KKLT model \cite{Kachru:2003aw}.

Some of the moduli, however, may remain  light, and they will play an important role in cosmology.
Indeed, in many string models, there are
often ultralight axions due to the above shift symmetry \cite{Svrcek:2006yi, Arvanitaki:2009fg, Conlon:2006tq},
even when all the other moduli get masses; unless
the symmetry is broken by appropriate non-perturbative effects generated
in the low energy, those string theoretic axions stay massless. 
Furthermore, their real component partners tend to remain relatively light\footnote{
Because such moduli can be stabilized through the SUSY-breaking effect,
their masses are comparable to or much lighter than the gravitino mass.
}. 
Although it certainly depends on the details of the model such as the properties of compact geometry and brane configurations,
the presence of such light moduli and ultralight axions
may be a natural outcome of string theories.\footnote{Here and in what follows, we often call
the real component of $T_{\rm moduli}$  the modulus, while the axion refers to its imaginary component
in the same multiplet. }

Let us focus on the lightest modulus and consider its impact on cosmology.
Since the modulus is light, it is likely deviated from the low-energy minimum
during inflation. After inflation, the
modulus starts to oscillate about the potential minimum with a large amplitude, and 
dominates the energy density of the Universe.\footnote{
This is the case even if the Hubble parameter during inflation is smaller than the modulus mass,
as long as the inflaton mass is heavier than the modulus~\cite{Higaki:2012ba}.
}  If its mass is of order the weak scale, it typically
decays during big bang nucleosynthesis (BBN), thus altering the light element
abundances in contradiction with observations.   This is the notorious cosmological moduli
problem~\cite{Coughlan:1983ci}. 

If the modulus mass is heavier than  O(10)\,TeV, on the other hand, it decays before  BBN, and the cosmological moduli problem 
will be greatly relaxed. Such high-scale SUSY breaking is also
suggested from the recently discovered standard-model-like 
Higgs boson~\cite{:2012gk,:2012gu}.
The cosmology of such moduli crucially depends on their decay modes. For instance, it was pointed out in 
Refs.~\cite{moduli,Dine:2006ii,Endo:2006tf} that
 the modulus generically decays into gravitinos with a sizable branching fraction if kinematically allowed,
and  those gravitinos produce  lightest SUSY particles (LSPs), whose abundance easily 
exceeds the observed dark matter (DM) density. 
Importantly, it was recently found by Kamada and the present authors in Ref.~\cite{Higaki:2012ba}
that the modulus decays into a pair of its axionic superpartners in the context of the (moderately) 
LARGE volume scenario \cite{Balasubramanian:2005zx},
and the produced axions will behave as extra radiation since the axions are effectively massless. 

The presence of such additional relativistic particles increases the expansion rate of the Universe, which affects
the cosmic microwave background (CMB) as well as the BBN yield of light elements, especially $^4$He and D. 
The amount of the relativistic particles is expressed in terms of
 the effective number of light fermion species, $N_{\rm eff}$, and it is given by $N_{\rm eff} \approx 3(3.046)$ before(after)
 the electron-positron annihilation in the standard cosmology.
 Interestingly, there is accumulating evidence for the existence of additional relativistic degrees of freedom coined ``dark radiation".
The latest analysis using the CMB data (WMAP7~\cite{Komatsu:2010fb} and SPT~\cite{Dunkley:2010ge})
has  given $N_{\rm eff} = 3.86 \pm 0.42$  (1$\sigma$ C.L.)~\cite{Keisler:2011aw}.
Other recent analyses can be found in Refs.~\cite{Hou:2011ec,GonzalezMorales:2011ty,Hamann:2011ge,Archidiacono:2011gq,Hamann:2011hu}.
The $^4$He mass fraction $Y_p$ is sensitive to
the expansion rate of the Universe during the BBN epoch, although  it has somewhat checkered history since it is very difficult to estimate systematic errors for deriving
the primordial abundance from $^4$He observations~\cite{Olive:2004kq}. Nevertheless, it is interesting that
an excess of $Y_p$ at the $2 \sigma$ level, $Y_p = 0.2565 \pm 0.0010\, ({\rm stat})
\pm 0.0050\, ({\rm syst})$,  was reported in Ref.~\cite{Izotov:2010ca}, which can be understood in terms of the
effective number of neutrinos, $N_{\rm eff} = 3.68^{+0.80}_{-0.70}$ $(2 \sigma)$.
Interestingly, it was recently pointed out that the observed deuterium abundance D/H also
favors the presence of extra radiation~\cite{Hamann:2010bk,Nollett:2011xk}:
$N_{\rm eff} = 3.90 \pm 0.44$ $(1 \sigma)$,  was derived from the CMB and
D/H data~\cite{Nollett:2011xk}.
It is intriguing that the CMB data as well as the Helium and Deuterium abundance
favor the presence of dark radiation, $\Delta N_{\rm eff} \sim 1$, while they are
sensitive to the expansion rate of the Universe at vastly different times.

\vspace{3mm}

In this paper, we consider both sequestered and non-sequestered models of LARGE volume scenario (LVS) 
in the singular regime \cite{Blumenhagen:2009gk}
as concrete examples. In both cases,  the overall volume modulus decays into a pair of axions, which
can account for the dark radiation suggested  by recent observations mentioned above.
(See e.g. Refs.\cite{Ichikawa:2007jv, Jaeckel:2008fi,
Nakayama:2010vs, Kobayashi:2011hp, Hasenkamp:2011em, Jeong:2012hp, Choi:2012zn, Graf:2012hb} for other models.)
Hence, the generation of the dark radiation from the modulus decay is a natural and robust prediction
in the LVS models. In order to estimate the dark radiation abundance, 
we study the various decay modes of the overall volume modulus, and find that
the decay through the Giudice-Masiero (GM) term \cite{Giudice:1988yz}
gives a dominant contribution to the total decay rate,
while the other decay modes such as the one into (transverse) gauge bosons as well as the three-body decays into
one scalar plus two fermions through the Yukawa couplings~\cite{Endo:2006qk} are suppressed. 
In the sequestered LVS, we will show that the LSPs produced by the modulus decay can naturally explain the observed DM abundance. 
In the non-sequestered case,  the QCD axion is a plausible candidate for DM. 

\vspace{3mm}

The rest of this paper is organized as follows. In Sec.~\ref{sec:2}, we give a
sketch of the modulus decay in LVS with an approximate no-scale structure. 
In Sec.~\ref{sec:3} we consider the sequestered LVS and estimate the abundance 
of dark radiation and dark matter in detail. We study the case of non-sequestered 
LVS in Sec.~\ref{sec:4}. The last section is devoted to discussion and conclusions.

\section{Preliminaries}
\label{sec:2}
Before proceeding to the realistic set-up, let us here summarize its important implications,
which can be easily understood as follows. In the following we adopt the Planck 
unit: $M_{\rm pl} \simeq 2.4 \times 10^{18}$ GeV $\equiv 1$.

The salient feature of the sequestered LVS model is an approximate no-scale structure of the lightest modulus $T$
which develops exponentially large vacuum expectation value (VEV); we consider the following 
K\"ahler and super-potentials and a constant gauge kinetic function as a low-energy effective theory,
\bea
\label{simpleK}
K &=& -3 \log\left[T+T^\dag - \frac{1}{3}\left\{\sum_i |Q_i|^2 + (z H_u H_d+{\rm h.c.})\right\} \right] + {\cal O}\left(\frac{1}{{\cal V}} \right), \non\\
& = &  -3\log(T+T^{\dag}) + \sum_i \frac{|Q_i|^2}{(T+T^{\dag})} + \left(z\frac{H_u H_d}{(T+T^{\dag})} +{\rm h.c.} \right) + \cdots , \quad \\
W&=& {\rm const}. + W_{\rm matter}(Q_i), 
\label{simpleW} 
\\
f &=& {\rm const}.
\eea
where $Q_i$ collectively denotes the matter fields including the Higgs field, and  $z = {\cal O}(1)$ is a numerical coefficient of the Giudice-Masiero (GM) term.
Note that the correction to the no-scale structure is  of order
the inverse of the large volume of the extra dimension ${\cal V} \sim (T+T^{\dag})^{3/2} \gg 1$,
which contains the effect of heavier moduli as well as ${\cal O}(\alpha'^3)$ corrections.
Because of the large VEV of $T$, the no-scale structure is protected from any corrections, especially 
from those violating the shift symmetry $T \to T + i \cdot ({\rm const})$. Thus, the axion $\sigma = {\rm Im(T)}$ is effectively massless. 

As shown in Ref.~\cite{Higaki:2012ba}, the modulus $\tau = {\rm Re}(T)$ decays into a pair of axions
through its kinetic term:
\beq
{\cal L} \supset - K_{T {\bar T}}\, \partial T^\dag \partial T \supset 
\frac{\delta \hat{\tau}}{\sqrt{6}} (\partial \hat{\sigma})^2 .
\eeq
Here $\delta \hat{\tau}$ and $\hat{\sigma}$ are the canonically normalized modulus and axion expanded around its VEV,
respectively.
The decay rate for this process is not suppressed by the volume, and therefore the axion production will be an important
decay process of the modulus. Similar process has been studied in the context of the non-thermal 
production of QCD axions from the saxion decay as an explanation of 
dark radiation~\cite{Ichikawa:2007jv, Hasenkamp:2011em, Jeong:2012hp, Choi:2012zn, Graf:2012hb}.

In addition to the decay into axions, the modulus $\tau$ decays mainly into the Higgs boson through the GM term. 
This can be understood as follows. The no-scale structure is unchanged up to an accuracy of our interest,
even if we redefine the modulus as
\beq
T' \;\equiv\; T-\frac{1}{3} z H_u H_d.
\eeq
Then the kinetic term of $T'$ contains a derivative coupling between $T$ and $H_uH_d$ as
\beq
{\cal L} \supset - K_{T' {\bar T'}}\, \partial T'^\dag \partial T' \supset 
\frac{z}{\sqrt{6}} (\partial^2 \delta \hat{\tau}) \hat{H}_u \hat{H}_d +{\rm h.c.}.
\eeq
Here $\hat{H}_{u,d}$ are the canonically normalized Higgses and we have performed integration by parts.
One can see that the decay rate into Higgs bosons is comparable to that into axions. This is also
expected to some extent because both arise from the kinetic term of $T$ (or $T'$).
It is also clear from this argument that 
the decay into Higgsino should be suppressed compared to that into Higgs bosons:
In terms of $T'$, there is no $H_u H_d$ dependence in the no-scale model, and so,
the higgsino bilinear term should appear only from $F^{T'}$ when expressed in terms of $T$ and $H_u H_d$.
However, since the scalar potential vanishes because of the no-scale structure,
no Higgs bilinear term appears at the leading order, and the higgsino mass as well as the modulus 
coupling to the Higgs bilinear should be suppressed. 

The other decay modes of the modulus can be most easily understood in terms of the
mixing with the conformal compensator field $\Phi$, whose interactions are given by
\beq
{\cal L}\;=\; \int d^4 \theta\, \Phi^\dag \Phi\left(-3  e^{-K/3} \right)  + \int d^2\, \theta \Phi^3 W + {\rm h.c.},
\eeq
where we consider the flat gravitational background. One can see from \REF{simpleK} that $T$ is mixed with $\Phi$,
and that the dependence of the compensator disappears in the scale invariant part of the interactions by an 
appropriate rescaling, $Q_i \Phi \rightarrow Q_i$. For instance, there will be no $\Phi$ dependence in the Yukawa coupling (and
also gauge interaction) after the rescaling. This explains why the three-body of $\tau$ into one scalar plus two fermions
through the Yukawa coupling is suppressed \cite{Endo:2006xg}. In a similar way, the decay into gauge bosons is suppressed at tree level.
In addition, we may ascribe the reason why the modulus $\tau$ decays into Higgs bosons at an unsuppressed rate
to the fact that  the GM term cannot absorb the compensator because of its holomorphic structure.

On the other hand, as mentioned above, the higgsino mass $\mu$ vanishes because
contributions of order ${\cal O(V}^{-1})$ are canceled in the no-scale
limit. Since the no-scale structure is broken at the next order in the volume expansion, 
non-zero $\mu$-parameter arises at order ${\cal O(V}^{-2})$. 
Furthermore, since the scale invariance is violated
at the quantum level, the modulus should be able to decay into (transverse) gauge bosons 
through the superconformal anomaly \cite{Dixon:1990pc}. 
Similar argument was used to understand the gravitino production from the inflaton decay in 
Refs.~\cite{Endo:2006xg,Endo:2007ih,Endo:2007sz}.
Finally, note that the anomaly mediation \cite{Randall:1998uk}
is sub-dominant effect because of the approximate no-scale structure.

So far, we have assumed the low-energy effective theory of the sequestered LVS. 
In the case of non-sequestered LVS, on the other hand, some of the features outlined
above do not apply. As we shall see, however, since the soft masses are much heavier than the
overall volume modulus in this case, the relevant decay modes are limited to those into the  
light higgs, longitudinal modes of $ZZ$ and $WW$ (the Nambu-Goldstone (NG) modes in the two Higgs multiplets) 
and axions, which greatly simplifies the argument. 

\section{The Sequestered LVS}
\label{sec:3}

Here let us study moduli stabilization in a local model with visible branes sitting
on a singularity~\cite{Blumenhagen:2009gk} within the type IIB orientifold compactifications.
The branes on the singularity lead to a chiral theory with anomalous $U(1)$ symmetries.
(For model building, see e.g. \cite{Aldazabal:2000sa}.)
In the following  we will show explicitly that the dark radiation composed of ultralight axions can be generated 
from the lightest modulus decay. 
In addition, the right amount of neutralino dark matter can be produced by the decay.

We consider the following K\"ahler potential and superpotentials\footnote{
We have neglected complex structure moduli 
dependence in the model for simplicity since they are irrelevant in this paper.
}:
\begin{align}
K &= -\log(S+S^{\dag}) 
- 2\log\bigg({\cal V} + \frac{\xi(S+S^{\dag})^{3/2}}{2} \bigg) 
+ \frac{(\tau_2 +\delta_{\rm GS} V_{U(1)})^2}{{\cal V}} + K_{\rm matter},
\label{EffLag1}
\\
{\cal V} &= \tau_1^{3/2}-\tau_3^{3/2}, \qquad
K_{\rm matter} = Z {|Q|^2} + Z_{\rm GM}H_u H_d + h.c. , 
\label{EffLag2}
\\
W &= W_{\rm flux} + A e^{-aT_3} +W_{\rm MSSM}(Q),
\label{EffLag3}
\\
f_{{\rm vis},a}& = \frac{1}{4\pi} (S+ \kappa_a T_2 ).
\label{EffLag4}
\end{align}
Here $S= e^{-\phi} + i C_0^{RR}$, where $\langle e^{\phi} \rangle = g_s$, 
is the string dilaton which is to be stabilized by $W_{\rm flux}$,
$T_i = \tau_i +i \sigma_i~(i=1,2,3)$ are K\"ahler moduli
and $V_{U(1)}$ denotes the anomalous $U(1)$ multiplet on the branes sitting at the singularity.
$\delta_{\rm GS}$ is a coefficient related with anomaly cancellation via the Green-Schwarz mechanism by the shift of $T_2$
under the anomalous $U(1)$.
For instance, $\delta_{\rm GS}$ is given by the mixed anomaly between $U(1)$ and $G_a$ in the MSSM gauge group, 
$\kappa_a \delta_{\rm GS} = \sum_i  q_i {\rm Tr}(T_a^2(\Phi_i))/\pi ={\cal O}(1/2\pi)$, 
where ${\rm Tr}(T_a^2(Q_i))$ is the dynkin index of
$\{Q_i\}$ which have charge $q_i$ under $U(1)$.
$\xi$ is given by a relation $\xi = -\chi({\rm CY})\zeta(3)/(4\sqrt{2}(2\pi)^3)$ coming from ${\cal O}(\alpha'^3)$ correction 
$\sim \int_{10~{\rm dim.}} {\cal R}^4$ and the consequent dilaton gradient \cite{Becker:2002nn}, 
and we will assume $\chi < 0$. 
$W_{\rm flux}$, which does not depend on the K\"ahler moduli, arises from closed string fluxes and the VEV is assumed to be of 
${\cal O}(1)$, 
and hence
this model will be natural from the view point of flux vacua landscape \cite{Giryavets:2003vd}.
The non-perturbative term originates from, e.g., the instanton brane ($O(1)$ E3-instanton) \cite{Blumenhagen:2009qh}
or $SO(8)$ gaugino condensation in the pure super Yang-Mills (SYM) sector 
on a stack of four D7-branes sitting on the top of an orientifold plane for RR-tadpole cancellation\footnote{
Note that the definition of the gauge coupling modulus on an orientifolded singularity 
receives a quantum correction discussed in \cite{Conlon:2009qa, Conlon:2010ji, Choi:2010gm}.
However, the redefinition of $T_3$ is irrelevant
and hence we will not consider the correction here.}.
Then one finds $a = 2\pi $ for $O(1)$ instanton or $\pi/3$ for pure $SO(8)$ SYM while
$A$ is expected to be of order unity.
Here we have assumed that the cycle related with $T_3$ modulus is a rigid one, 
where the adjoint fields, i.e. open string moduli, are absent.

For the matter sector, we consider the minimal supersymmetric Standard Model (MSSM) 
localized on the singularity whose ``volume" is described by $T_2$.
The coefficient $\kappa_a$ in the gauge coupling depends on the structure of the singularity and 
that of the dilaton in the gauge coupling is set to be unity
for simplicity.
$Z$ is the matter K\"ahler metric and we have introduced Giudice-Masiero (GM) term $Z_{\rm GM}$.
The latter GM term is very important not only for generation of  the higgsino mass 
but also for obtaining a right amount of dark radiation.

\subsection{The moduli stabilization}

We then find the (meta-)stable vacuum in the
the scalar potential
\begin{align}
V= e^{K}\left[|DW|^2-3|W|^2 \right] + \frac{1}{2}D^2 
\end{align}
through the moduli stabilization:
\begin{align}
\langle \tau_3 \rangle \sim  \frac{\xi^{2/3}}{g_s}, \qquad
{\cal V} \sim \langle \tau_1^{3/2} \rangle \sim \frac{W_0}{aA}\sqrt{\langle \tau_3 \rangle}
e^{a \langle \tau_3 \rangle},
\qquad 
\langle \tau_2 \rangle = 0.
\end{align}
Here $T_3$ is stabilized almost supersymmetrically a la KKLT with 
$F^{T_3} \sim 1/ (\log({\cal V}){\cal V}) \lesssim 1/{\cal V}$ \cite{Kachru:2003aw}
and $T_1$ is fixed by the SUSY-breaking effect as a result of the competition between
$\xi$ and non-perturbative terms,
while $T_2$ is done via the D-flat SUSY condition, $D_{U(1)} \sim \partial_{T_2}K =0$ and it is then eaten by the $V_{U(1)}$ gauge multiplet.
Note that axions except for $\sigma_1$ will be stabilized at the origin, while $\sigma_1$ remains massless.
For the visible coupling, the dilaton can be stabilized as $\langle S \rangle \simeq 24$ (in the MSSM with TeV scale soft masses).
Here note that the vacuum considered so far has the negative cosmological constant with broken SUSY, 
and hence it is necessary to add the uplifting potential for obtaining the de Sitter/Minkowski vacuum.
For the uplifting potential, one can consider an explicit SUSY breaking term originating from the sequestered anti-D3-brane 
on the tip of the warped throat \cite{Klebanov:2000hb, Giddings:2001yu}:
\begin{align}
\delta V \equiv  V_{\rm uplift}= \epsilon e^{2K_{\rm moduli}/3} \sim \frac{\epsilon}{{\cal V}^{4/3}}
,~~~{\rm where}~
\epsilon \sim \frac{1}{\log({\cal V}){\cal V}^{5/3}} .
\end{align}
The factor of $\epsilon$ is the minimum of the warp factor.
For another option, dynamical SUSY breaking models with non-zero $F$ or $D$-term VEVs are also possible \cite{Cicoli:2012fh}.
(See also \cite{Saltman:2004sn, Burgess:2003ic}.) 
In order to realize the SUSY-breaking soft mass of order TeV or so, 
we will consider the case
\begin{align}
{\cal V} \sim {\cal O}(10^7) 
\end{align}
by choosing a proper manifold such as the value of $\xi \propto -\chi({\rm CY})$ is positive.

One can express the moduli fields in terms of the mass eigenstates $(\delta \phi, \delta a)$
 as follows \cite{Cicoli:2010ha}:
\begin{align}
\begin{pmatrix}
\delta \tau_1 \\
\delta \tau_2 \\
\delta \tau_3 \\
\delta (e^{-\phi}) 
\end{pmatrix}
\sim 
\begin{pmatrix}
{\cal V}^{2/3} \\
0 \\
\frac{m_{3/2}}{m_{\tau_3}} \\
{\cal V}^{-1/2}
\end{pmatrix}
\delta \phi_1 +
\begin{pmatrix}
0 \\
{\cal V}^{1/2} \\
0 \\
0
\end{pmatrix}
\delta \phi_2 +
\begin{pmatrix}
{\cal V}^{1/6} \\
0 \\
{\cal V}^{1/2} \\
{\cal V}^{-1}
\end{pmatrix}
\delta \phi_3 +
\begin{pmatrix}
{\cal V}^{1/6} \\
0 \\
{\cal V}^{-1/2} \\
{\cal O}(1)
\end{pmatrix}
\delta \phi_s ,
\\
\begin{pmatrix}
\delta \sigma_1 \\
\delta \sigma_2 \\
\delta \sigma_3 \\
\delta C_0^{\rm RR} 
\end{pmatrix}
\sim 
\begin{pmatrix}
{\cal V}^{2/3} \\
0 \\
0 \\
0
\end{pmatrix}
\delta a_1 +
\begin{pmatrix}
0 \\
{\cal V}^{1/2} \\
0 \\
0
\end{pmatrix}
\delta a_2 +
\begin{pmatrix}
{\cal V}^{1/6} \\
0 \\
{\cal V}^{1/2} \\
{\cal V}^{-1}
\end{pmatrix}
\delta a_3 +
\begin{pmatrix}
0 \\
0 \\
{\cal V}^{-1/2} \\
{\cal O}(1)
\end{pmatrix}
\delta a_s .
\end{align}
Note that $\delta \tau_2$ and $\delta \sigma_2$ do not mix with the other fields because of $\tau_2 = 0$ and
$a_1$ is decoupled from the MSSM sector localized on the $T_2$-cycle.
Then, one obtains
\begin{align}
m_{3/2} &= e^{K/2}W \sim m_S \sim \frac{1}{{\cal V}}, \\
m_{\phi_1} &\sim \frac{1}{\sqrt{\log(V)}{\cal V}^{3/2}} , \quad
m_{\phi_2,a_2} = m_{V_{U(1)}} \sim \frac{1}{{\cal V}^{1/2}} , \quad
m_{\phi_3,a_3} \sim \frac{\log({\cal V})}{{\cal V}} .
\end{align}

The resultant SUSY-breaking $F$-terms are given by
\begin{align}
\frac{F^{T_1}}{2\tau_1} \sim \frac{1}{{\cal V}} \left(
1 + 
{\cal O}
\left(
\frac{1}{\log({\cal V}) {\cal V}}
\right)
\right),~~~
F^{T_2} \propto \tau_2 = 0, ~~~
\frac{F^{T_3}}{2\tau_3} \sim \frac{1}{\log({\cal V}){\cal V}}.
\end{align}
In addition to these, the non-zero $F^S$ is obtained and will be expected to be
\begin{align}
\frac{F^S}{S+S^{\dag}} &\sim
\frac{1}{{\cal V}}  \bigg( - \frac{W}{S+S^{\dag}} + \partial_S W_{\rm flux} + \frac{\xi}{{\cal V}} W \bigg) \\
& \sim \frac{1}{\log({\cal V}){\cal V}^2} = {\cal O}\bigg( \frac{1}{{\cal V}^2}\bigg)
\end{align}
due to the $\xi$-dependent SUSY-breaking term.
This is because it is expected that without the $\alpha'$-correction  $\xi$, the dilaton would be stabilized supersymmetrically,
i.e., $- {W}/{(S+S^{\dag})} + \partial_S W_{\rm flux} = 0$ in the supersymmetric flux backgrounds. 
In addition, there would be a small cancellation in the vacuum value of $F^S$, although this  depends on the choice of the fluxes.
Hence, $1/\log({\cal V}) ={\cal O}(0.1)$ implies such  small cancellation.
Note that the compensator $F$-term is similarly suppressed
\begin{align}
F^{\Phi} = m_{3/2} + \frac{\partial_I K}{3}F^{I} \sim \frac{1}{\log({\cal V}){\cal V}^2}.
\end{align}
Hence, because of the no-scale structure, 
the anomaly mediation has only sub-dominant effects on the soft masses, 
and its size is suppressed by a loop factor compared to the modulus mediation.

Finally, we would like to give a comment on possible corrections to the axion mass.
Even if there is an instanton correction wrapping on the big cycle, 
$\delta K \sim k(T_1 +T_1^{\dag})e^{-2\pi T_1} + h.c.$ or
$\delta W = A_b e^{-2\pi T_1}$, 
the following discussion will not change due to the large volume
of extra dimension:
\begin{align}
\delta m_{a_1} \sim e^{-2\pi {\cal V}^{2/3}} \sim 10^{-10^5} {\rm eV} \ll 0.1 {\rm eV}.
\end{align}
Thus, the presence of the ultralight axion is plausible and hence the axion becomes dark radiation 
as seen below.

\subsection{Matter sector and soft SUSY breaking parameters}

Next, let us estimate the soft SUSY-breaking terms in the visible sector on the singular $T_2$-cycle.
The gaugino masses are given by the dilaton F-term:
\begin{align}
M_{1/2} &\sim \frac{F^S}{S+S^{\dag}} \sim \frac{1}{\log({\cal V}){\cal V}^2}. 
\end{align}
Hence the other soft-SUSY breaking terms are expected to appear (at least) at this order;
this implies that the locality of the local sector is broken down at this order in the large volume expansion.

With respect to $A$-terms and scalar masses, the precise information of $Z$ is required.
Though the correct metric is unknown,
it is expected that the locality of the local brane sector enables us to calculate the soft masses \cite{Blumenhagen:2009gk}.
Once one requires that local Yukawa couplings 
are dependent just on the local modulus at the leading order of perturbation by string coupling 
or by $\alpha'$-correction \cite{Conlon:2011jq},
\begin{align}
\frac{e^{K_{\rm moduli}/2}Y}{\sqrt{Z^3}} \sim \frac{Y}{(\tau_1^{3/2}\sqrt{Z^3})} \equiv y(\tau_2) + ({\rm tiny~corrections}),
\end{align}
one finds
\begin{align}
Z \sim e^{K_{\rm moduli}/3} \qquad {\rm or} \qquad Z \sim \frac{1}{\tau_1} .
\end{align}
Here tiny corrections, which would be of ${\cal O}({\cal V}^{-2})$ at least, can include K\"ahler moduli other than $T_2$.

Suppose that there is an approximate shift symmetry in the Higgs sector, $H_{u,d} \to H_{u,d} + i \cdot({\rm const.})$,
which might come from the higher dimensional gauge fields \cite{Hebecker:2012qp}.
If this symmetry exactly holds, the K\"ahler potential in the Higgs sector is given by
\begin{align}
K_{\rm Higgs} = Z|H_u + H_d^{\dag}|^2  + {\cal O}(|H_u + H_d^{\dag}|^4) ,
\end{align}
implementing $Z=Z_{\rm GM}$.
Thus one can write the GM term as
\begin{align}
Z_{\rm GM} \sim z e^{K_{\rm moduli}/3} \qquad {\rm or} \qquad Z_{\rm GM} \sim \frac{z}{\tau_1} .
\label{ZGM}
\end{align}
Here $z$ is a constant and therefore $z-1$ represents a possible breaking of the shift symmetry by the compactification.
Alternatively, it is possible to change  the effective value of $z$ without violation of the shift symmetry. 
Let us extend the Higgs sector and introduce $n$ additional pairs of the Higgs doublets $(H_u^\prime, H_d^\prime)$
which satisfy the shift symmetry. Namely, all the Higgs doublets are assumed to have exactly the same
interaction in (\ref{ZGM}) with $z=1$. 
In string compactifications, such an extended Higgs sector can be naturally realized \cite{Higaki:2005ie}.
Then, the decay rate into these Higgs bosons will be enhanced by a factor of $(1+n)$,
which effectively corresponds to the case of $z = \sqrt{1+n}$.  As we shall see later, the right amount of dark radiation and
dark matter can be generated for $z \sim 1.5\pm0.3$, which can be nicely explained if $n=1$ or $2$.

In the similar way to the Yukawa coupling, when one demands that
a n($>3$)-point coupling in the superpotential is dependent just on the local modulus, 
the cutoff scale is found as \cite{Conlon:2009qa, Choi:2010gm}
\begin{align}
M_{\rm cutoff} \sim \sqrt{Z} M_{\rm pl} \sim \frac{M_{\rm pl}}{{\cal V}^{1/3}} .
\end{align}

Next, with the above K\"ahler potential we can read the remaining soft masses
\begin{align}
A_{i_1 \cdots i_n} &\sim \frac{1}{\log({\cal V}){\cal V}^2} - \frac{1}{{\cal V}^2} , \\
m_0^2 & \sim  \left(\frac{1}{\log({\cal V}){\cal V}^2}\right)^2 - \frac{1}{{\cal V}^3}, \\
\mu & \sim \frac{z}{\log({\cal V}){\cal V}^2}, \\
B\mu & \sim z\left(\frac{1}{\log({\cal V}){\cal V}^2}\right)^2 - \frac{z}{{\cal V}^3} .
\end{align}
What is important is that soft scalar mass can be suppressed compared to the lightest modulus mass:
\begin{align}
m_0 \lesssim m_{\tau_1} \sim \frac{1}{\sqrt{\log({\cal V})}{\cal V}^{3/2}}.
\end{align}
For the concreteness, we will take
\begin{align}
m_0 \sim \frac{1}{{\cal V}^{2}} = {\cal O}(10^3- 10^4){\rm GeV},
\label{m0eg}
\end{align}
and then
\begin{align}
\label{m12eg}
M_{1/2} &\sim A \sim 
\frac{1}{\log({\cal V}){\cal V}^{2}} ={\cal O}(10^2-10^3) {\rm GeV}, \\
\label{mueg}
\mu &\sim \frac{z}{\log({\cal V}){\cal V}^2} ={\cal O}(10^2-10^3) {\rm GeV}, \\
B\mu &\sim z m_0^2 \sim
\frac{z}{{\cal V}^4} = {\cal O}(10^6- 10^8){\rm GeV}^2.
\end{align}
Such heavy scalars are compatible with the standard-model like
Higgs boson of mass about $125$\,GeV.
For the moduli sector one will find then
\begin{align}
m_{\phi_1} &= {\cal O}(10^{6} -10^{7}){\rm GeV}, \quad
m_{\phi_2, a_2} = {\cal O}(10^{15}){\rm GeV}, \quad
m_{\phi_3, a_3} = {\cal O}(10^{12}){\rm GeV}, \\
m_{3/2} &= {\cal O}(10^{10}- 10^{11}){\rm GeV}.
\end{align}
Note that the decay of lightest modulus to the gravitinos will be kinematically forbidden.

\subsection{Decay modes of the lightest modulus: Generation of dark radiation}

We will consider the decay mode of the lightest modulus $\delta \phi_1$, 
assuming that the coherent oscillation of the lightest modulus $\delta \phi_1$ is induced after inflation and the
energy density of the modulus dominates the Universe some time after the inflaton decay but before the modulus decay. 
The Hubble scale during inflation should satisfy $H_{\rm inf} \lesssim m_{\phi_1}$ for evading a problem of
run-away and decompactification.
Note that the cosmological moduli problem is not solved even for such low-scale inflation,
if the inflaton mass is larger than the modulus mass
(See also discussion on moduli problem in Sec. III B of \cite{Higaki:2012ba}).

For the SUSY breaking parameters given in the previous subsection (especially $m_0 \lesssim m_{\phi_1}$),
there are two main decay modes, $\delta \phi_1 \to H_u H_d$ and $\delta \phi_1 \to 2a_1$. The interaction of the modulus
to Higgs arises from the GM term:
\begin{align}
{\cal L} \supset \int d^4 \theta \frac{z H_u H_d}{(T_1 + T_1^{\dag})}+{\rm h.c.} 
\supset \frac{ z}{\sqrt{6} M_{\rm pl}} (\partial^2 \delta \phi_1)
\left( \hat{H}_u {\hat H}_d + {\rm h.c.} \right)
\end{align}
where $\hat{H}_u$ and $\hat{H}_d$ are canonically normalized up- and down-type Higgs bosons.
The partial decay rate for the modulus decay into Higgs bosons  is given by
\bea
\Gamma(\delta \phi_1 \to H_u H_d) &\simeq&  \frac{2z^2}{48 \pi} \frac{m_{\phi_1}^3}{M_{\rm pl}^2}.
\eea

On the other hand, the coupling of the modulus to the axions arises from the kinetic term~\cite{Higaki:2012ba},
\begin{align}
{\cal L} \supset  \frac{2}{\sqrt{6} M_{\rm pl}} \delta \phi_1\partial_{\mu} a_1 \partial^{\mu}a_1.
\end{align}
The partial decay rate for the modulus decay into axions is 
\bea
\Gamma(\delta \phi_1 \to 2a_1) &\simeq& \frac{1}{48\pi}  \frac{m_{\phi_1}^3}{M_{\rm pl}^2}.
\label{Gam2a}
\eea
Therefore the total decay rate is approximately given by
\bea
\Gamma_{\delta \phi_1} &\simeq& \Gamma(\delta \phi_1 \to  H_u H_d) + \Gamma(\delta \phi_1 \to 2a_1) \\
& \simeq& \frac{2z^2+1}{48 \pi} \,\frac{m_{\phi_1}^3}{M_{\rm pl}^2} 
\label{maindecay}
\eea
and the branching fraction to the axions is given by
\begin{align}
B_a \equiv B(\delta \phi_1 \to 2a_1) \simeq \frac{1}{2z^2 + 1 }.
\end{align}
It should be emphasized, the branching fraction is generically of order ${\cal O}(0.1)$ for $z = {\cal O}(1)$.
Thus, the presence of dark radiation is a robust prediction of this scenario. 

Taking account of the axion production, the reheating temperature of the modulus is estimated as
\begin{align}
\label{Td}
T_d &\simeq (1-B_a)^\frac{1}{4} \left(\frac{90}{\pi^2 g_*(T_d)}\right)^{1/4}
\sqrt{\Gamma_{\delta \phi_1} M_{\rm pl}}  \\
&\simeq
1.7\, {\rm GeV} \times  (1-B_a)^\frac{1}{4} \left(\frac{2z^2+1}{3}\right)^{1/2}
\left(\frac{80}{g_*(T_d)}\right)^{1/4} 
\left(\frac{m_{\phi_1}}{10^7{\rm GeV}}\right)^{3/2}.
\end{align}
Thus, using the expression of $\Delta N_{\rm eff}$ in terms of $B_a$ and $T_d$ given in
Appendix \ref{dNeff}, one obtains
\begin{align}
\Delta N_{\rm eff} \simeq \frac{43}{14z^2}\times \lrfp{10.75}{g_*(T_d)}{\frac{1}{3}} .
\end{align}
If we substitute $z=1.2$ and $g_* (T_d)=80$, we obtain $\Delta N_{\rm eff} \approx 1.1$.

\begin{figure}[t!]
\begin{center}
\includegraphics[width=7cm]{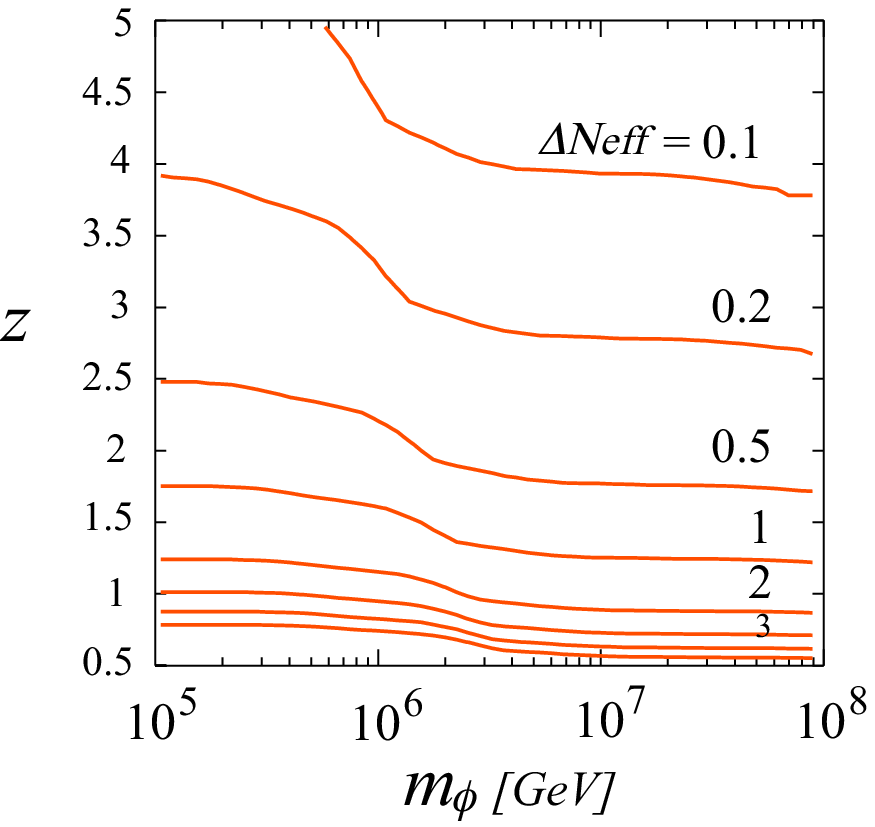} \qquad
\includegraphics[width=7cm]{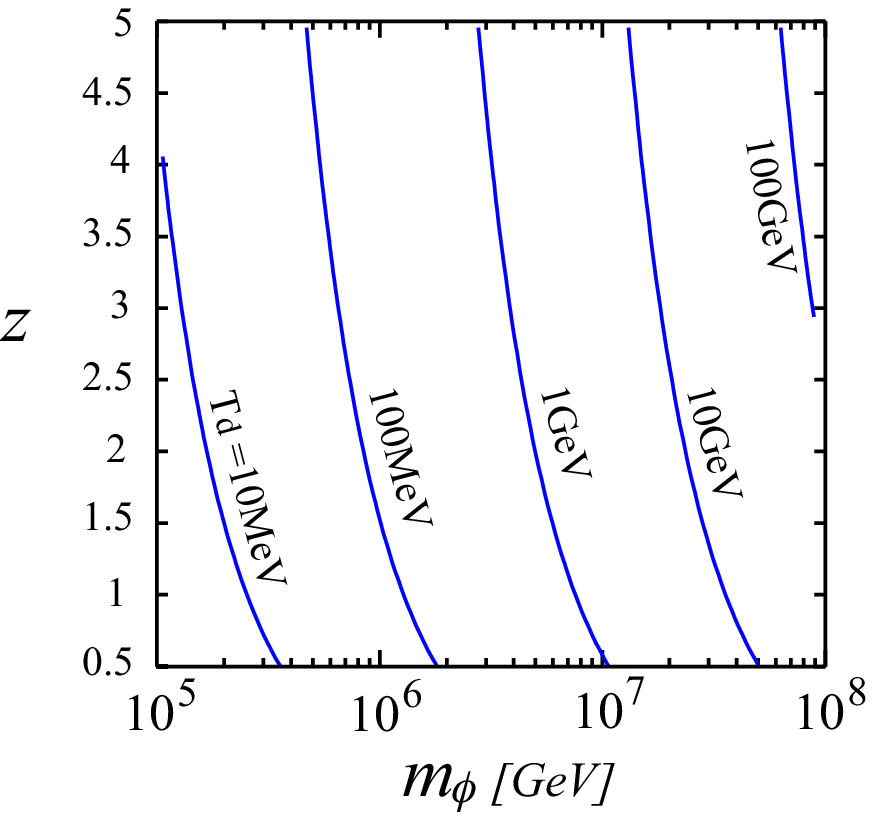} \qquad 
\caption{Contours of the additional effective number of neutrinos, $\Delta N_{\rm eff}$, the modulus reheating temperature, $T_d$
in the plane of the lightest modulus mass $(m_\phi)$ and the coefficient of the GM term $z$.
Here 
the higgsino mass is given by the relation $\mu \simeq \frac{z}{{\log({\cal V})^{1/3}}} (m^{4}_{\phi}/M_{\rm pl})^{1/3}
\approx  \frac{z}{2.55} (m^{4}_{\phi}/M_{\rm pl})^{1/3} $ in the above two figures.}
\label{fig:dN}
\end{center}
\end{figure}

In Fig.~\ref{fig:dN}, we show the contours of the additional effective number of neutrinos, $\Delta N_{\rm eff}$ and
the modulus reheating temperature, $T_d$ in the plane of the modulus mass $(m_\phi)$ 
and the coefficient of the GM term ($z$). Since the relativistic degrees of freedom $g_*$ sensitively depends on 
the reheating temperature~\cite{Laine:2006cp}, we have taken into account of the dependence, solving Eq.~(\ref{Td}) 
numerically by iteration. We can see that $z \sim 1.5\pm0.3 $ is needed in order to account for the dark radiation with
$\Delta N_{\rm eff} \sim 1$.

\subsubsection{The decay rate into another Standard Model particles: Sub-dominant decay rate}

Here we would like to give a comment on the decay rate of the lightest modulus
from the point view of no-scale model with the conformal compensator. 
The decay of $\delta \phi_1$ can be understood in terms of the mixing with the compensator field. See 
Refs.~\cite{Endo:2006xg,Endo:2007ih,Endo:2007sz}. The decay into two scalars, ${\tilde Q}_i$ and ${\tilde Q}_i^\dag$, 
can be easily seen to be suppressed by the soft mass. One may expect that such suppression can be avoided
if one attaches a Yukawa coupling to one of the external line of the scalars, so that the moduli
decays into one scalar (stop) and two fermions (higgsino and top), 
instead of the two scalars. This would be actually the case
for a generic form of the K\"ahler potential, however, for the K\"ahler potential of the no-scale structure,
such decay is also suppressed~\cite{Endo:2006xg}. This is because the Yukawa coupling is dimensionless, and therefore
the compensator field does not couple at the tree level. 

In a similar fashion, we can understand why the moduli decay into gauge sector is suppressed at tree level.
This is because the gauge interaction is scale invariant at the classical level. Of course, at the quantum level,
the scale invariance is violated, and this is nothing but the superconformal anomaly \cite{Dixon:1990pc}:
\begin{align}
\frac{1}{g^2(E)} &= {\rm Re}(f_a) + \frac{b_a}{16\pi^2}\log\left( 
{e^{K_{\rm moduli}/3}}
\right) + \cdots \\
& \supset {b_a}\times {\cal O}(10^{-2}) \times \frac{\delta \phi_1}{M_{\rm pl}}.
\end{align}
Thus one reads the modulus coupling to the gauge boson at the quantum level.
Here we omitted the correction from usual running and 
the matter K\"ahler potential which is proportional to $\log (e^{-K_{\rm moduli}/3}Z)$
since in the local (no-scale) model this term becomes irrelevant.
$b_a \equiv -3T_g + T_r = (-3, 1, 33/5)$ is the beta function coefficient of the $SU(3),~SU(2)$ and $U(1)$ gauge group
respectively,
$T_g$ and $T_r$ are the Dynkin index of the adjoint representation and matter fields in the
representation $r$, which are normalized to $N$ for $SU(N)$ and $1/2$ for its fundamentals, respectively.
Hence the modulus (compensator) is coupled also to the Yukawa coupling at the quantum level
through the running from the cutoff scale $M_{\rm pl}/{\cal V}^{1/3}$, too.

The decay rate of $\delta \phi_1$ into a pair of (transverse) gauge bosons is given by\footnote{
Note that a half of the decay rate given in Refs.~\cite{Endo:2007ih,Endo:2007sz} comes from the decay into gauginos, which
is suppressed in our case, and that the canonical normalization was assumed.
To summarize, the decay rate can be obtained by multiplying the rate in Eq.~(28) of Ref.~\cite{Endo:2007sz} with $1/(2 K_{T {\bar T}})$.
}
\beq
\Gamma(\delta \phi_1 \rightarrow 2 A_\mu)\;\simeq\; \frac{N_g \alpha^2 b_a^2}{4608 \pi^3} \frac{m_{\phi_1}^3}{M_p^2},
\label{decayintogauge}
\eeq
where $N_g = (8,3,1)$ is the number of the gauge bosons of the corresponding gauge symmetry $SU(3),~SU(2)$ and $U(1)$ 
in the MSSM respectively, and
$\alpha = g^2/4\pi$ denotes the gauge coupling. 
Hence this is a sub-dominant fraction of the modulus decay rate.

\subsubsection{SUSY particle production rate: Generation of neutralino dark matter}

Since the lightest moduli is stabilized in a non-supersymmetric fashion, 
the decay into gauginos and higgsinos is suppressed compared to that into gauge bosons and Higgs~\footnote{
This is not  the case in the geometric regime where the gauge kinetic function depends
on the another type of K\"ahler modulus which mixes with the overall volume modulus~\cite{Higaki:2012ba}. 
}.
Through the couplings
\begin{align}
{\cal L} \supset  \frac{M_{1/2}}{M_{\rm pl}} \delta\phi_1 \lambda \lambda + 
\frac{\mu}{M_{\rm pl}} \delta\phi_1 \tilde{h} \tilde{h} ,
\end{align}
one finds that the branching fraction into gauginos and higgsinos
is suppressed by the volume, $\sim (\mu/m_{\phi_1})^2 \sim (M_{1/2}/m_{\phi_1})^2 \sim 1/{\cal V} \sim 10^{-7}$.

There are additional contributions of the same order or slightly larger, which do not depend on the $\mu$-term
or gaugino mass. For instance, we may consider a diagram of the modulus decaying into a pair of gluons
with one of the gluons splitting into squark and anti-squark. The rate of the diagram is considered to be suppressed
by a factor of $O(10^{-2})$ compared to (\ref{decayintogauge}) due to the 
phase space factor of three-body decay and the gauge coupling of the strong interaction. 
This gives the branching fraction of the squark production of order $10^{-6}$. 

Actually, however, the main source for the SUSY particle production comes from the following processes.
The heavy Higgs bosons produced from the modulus decay will decay into Higgsino and Wino/Bino,
if kinematically allowed. (This is the case for the adopted mass spectra (\ref{m0eg}), (\ref{m12eg}) and (\ref{mueg}).)
The branching fraction of this process will be of ${\cal O}(0.1)$. Even if this decay modes are kinematically forbidden,
an electroweak correction to the main decay mode, $\delta \phi_1 \rightarrow$ 2 Higgs bosons
will give a branching fraction  of order $10^{-3}$. In fact, as we shall see below, the dark matter abundance does
not depend on  the precise value of the branching fraction, and our results are valid for the branching
fraction greater than about $10^{-5}$.

The non-thermally produced LSPs will contribute to the dark matter abundance,
if the R-parity is conserved. The thermal relic abundance is negligible because the
reheating temperature is lower than the freeze-out temperature. In the following analysis
we assume that the dark matter consists of the single component, for simplicity. 
For the parameters of our interest, the produced LSPs annihilate effectively, and the
final LSP abundance is given by
\bea
\frac{n_{\chi}}{s} &\simeq& \left(\sqrt{\frac{8 \pi^2 g_*(T_d)}{45}} \langle \sigma v \rangle M_{\rm pl} T_d \right)^{-1},
\eea
where $n_\chi$ denotes the  number density of the LSP, and $\langle \sigma v \rangle$ is the thermally averaged annihilation.
In terms of the density parameter, it is given by
\beq
\Omega_\chi h^2 \;\simeq\; 0.16 \lrfp{g_*(T_d)}{80}{-\frac{1}{2}} \lrfp{\langle \sigma v \rangle}{3 \times 10^{-8} {\rm \,GeV}^{-2}}{-1}
\lrfp{T_d}{1\,{\rm GeV}}{-1} \lrf{m_\chi}{500{\rm \,GeV}}.
\label{OLSP}
\eeq
Thus, it is possible to account for the observed DM density by the LSPs non-thermally produced by the modulus decay,
if the annihilation cross section is relatively large as in the case of the Wino-like LSP.

\begin{figure}[t!]
\begin{center}
\includegraphics[width=8cm]{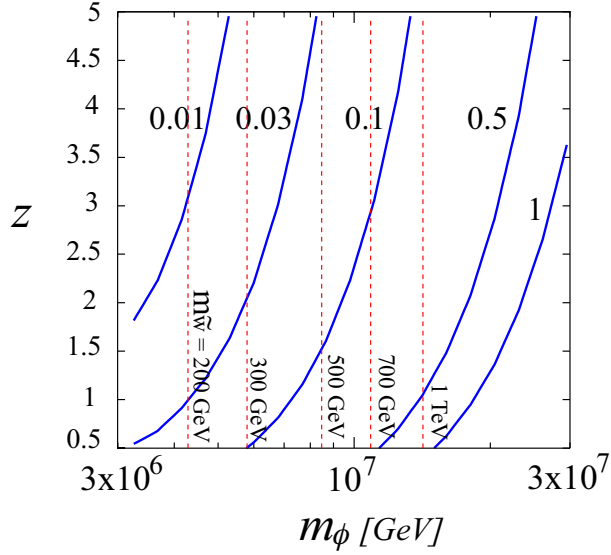} 
\caption{Contours of the non-thermally produced Wino dark matter abundance, $\Omega_\chi h^2 = 0.01, 0.03, 0.1, 0.5$ and $1$ (solid (blue) lines)
and the Wino mass,  $m_{\tilde W} = 200, 300, 500, 700,$ and $1000$\,GeV (dashed (red) lines), 
in the plane of the modulus mass $(m_\phi)$ and the coefficient of the GM term ($z$).
The relation $m_{\tilde W} = 1/(\log({\cal V}) {\cal V}^2)$ is assumed.
Here recall that $ \mu \sim z m_{\tilde{W}}$.
}
\label{fig:Odm}
\end{center}
\end{figure}

We have shown in Fig.~\ref{fig:Odm} the contours of the DM abundance as a function of $m_\phi$ and $z$.
Here we have assumed the Wino-like LSP with the mass $m_{\tilde W} = 1/(\log({\cal V}) {\cal V}^2)$.
The branching fraction of the SUSY particle production is set to be $0.1$ in this
figure, although the results is insensitive to the precise value of the branching fraction, because the final abundance is determined by
the annihilation (see (\ref{OLSP})).
We have numerically confirmed that our results remain almost intact for the branching fraction
between $10^{-5}$ and $0.1$.
We can see from the figure that the observed DM density, $\Omega_\chi h^2
= 0.1126 \pm 0.0036$~\cite{Komatsu:2010fb},  can be explained for 
$m_\phi = 6 \times 10^6{\rm\,GeV} - 10^7{\rm\,GeV}$ and the Wino mass $m_{\tilde W}
= 300 - 1000$\,GeV. 
Interestingly, if we require the right amount of dark radiation $\Delta N_{\rm eff} \sim 1$, which is realized
for $z \sim 1.5$, the Wino mass should be around $500$\,GeV. 

\begin{figure}[t!]
\begin{center}
\includegraphics[width=9cm]{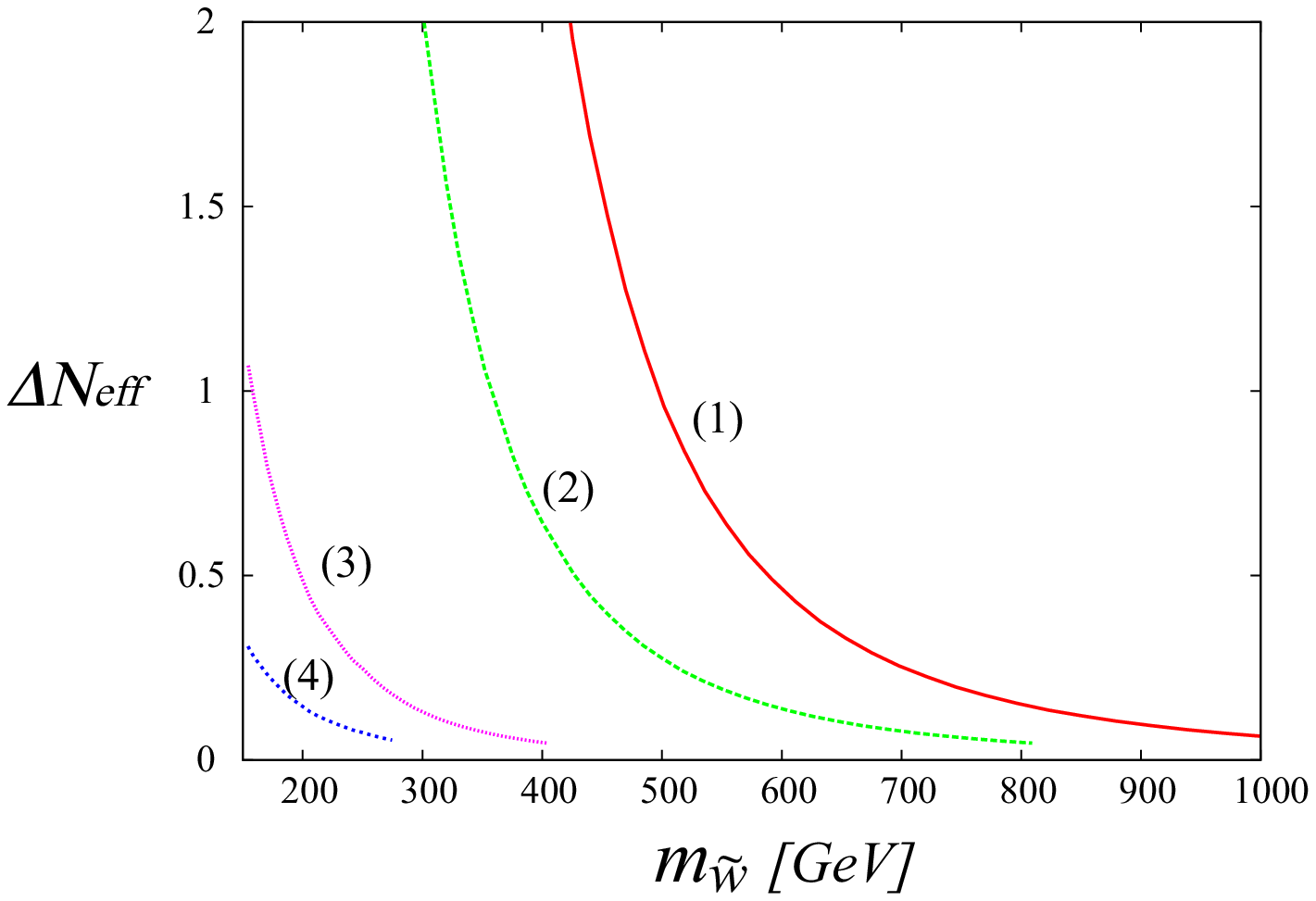} 
\caption{ The relations between $\Delta N_{\rm eff}$ and the Wino mass $m_{\tilde W}$ are shown for the following
four cases: 
(1) $m_{\tilde W} = 1/(\log({\cal V}) {\cal V}^2)$, (2) $m_{\tilde W} = 0.1/({\cal V}^2)$,
(3)  $m_{\tilde W} = 0.3/({\cal V}^2)$, and (4)  $m_{\tilde W} = 0.5/({\cal V}^2)$.
The right DM abundance,  $\Omega_\chi h^2 \simeq 0.11$, is imposed. 
}
\label{fig:dN-mwino}
\end{center}
\end{figure}

It is interesting that $\Delta N_{\rm eff}$ and the Wino mass
are related to each other when we impose $\Omega_\chi h^2 \simeq 0.11$.
In order to see how the relation changes, we have examined several cases of the Wino mass dependence
on ${\cal V}$.
See Fig.~\ref{fig:dN-mwino}. We can see that $\Delta N_{\rm eff}$ decreases with respect to the Wino mass. This can be
understood as follows. Since we impose $\Omega_\chi h^2 \simeq 0.11$, the reheating temperature and so $z$ needs to increase 
as the Wino mass, which in turn reduces the branching fraction of the axion production, thereby decreasing 
$\Delta N_{\rm eff}$.
In particular, if we adopt e.g. $m_{\tilde W} = 1/(2{\cal V}^2)$,
the dark radiation abundance tends to be small, $\Delta N_{\rm eff} \lesssim 0.3$.
We have numerically checked that the LSP abundance tends to be
too large if the LSP is the Higgsino-like neutralino.

\section{The non-sequestered LVS}
\label{sec:4}

In this section, we will study the non-sequestered LVS,
where the breakdown of no-scale structure in the visible sector is induced by the quantum effect 
and hence the soft terms become comparable to or smaller than the gravitino mass.
Such loop corrections will be obtained when the visible branes are sitting on an orientifolded singularity
or an orbifold singularity with D3 and D7-branes \cite{Conlon:2009qa, Conlon:2010ji, Choi:2010gm}
and when the threshold correction from the massive gauge boson to the soft terms
\cite{Shin:2011uk} is not negligible in LVS.
(See also \cite{Berg:2007wt} for the discussion of 1-loop K\"ahler potential of moduli.)
If we require the SUSY breaking at TeV - PeV in the visible sector,  the volume ${\cal V}$ should be about
$10^{12 - 15}$, and so,  $\delta \phi_1$ has a mass lighter than $\sim 1$\,GeV. Thus, the cosmological moduli problem revives.
In order to avoid the serious moduli problem, we assume that the soft masses are at an intermediate scale so that
the lightest modulus decays before BBN. The purpose of this section is
to see that the ultralight axion can again account for the observed amount of the dark radiation 
even for this case.
Hence the dark radiation is a robust prediction
of our framework.

Let us study the model in more detail. 
The effective model is given as Eq.(\ref{EffLag1}), (\ref{EffLag2}), (\ref{EffLag3}) and (\ref{EffLag4}),
being replaced with redefined moduli $\tilde{\tau}_2$ and $\tilde{\tau}_3$
\begin{align}
\tilde{\tau}_{i} = \tau_i - \frac{1}{3}\alpha_{i} \log({\cal V}), \qquad
{\alpha_i} = {\cal O}\bigg(\frac{1}{2\pi}\bigg) \qquad
(i=2,3).
\end{align}
This is because moduli $\tau_i$ corresponds to the inverse of the gauge coupling $4\pi/g^2$,
which will start to run not from the winding scale $M_{\rm pl}/{\cal V}^{1/3}$ 
but from the string scale $M_{\rm pl}/{\cal V}^{1/2}= {\cal O}(10^{14}-10^{15})$ GeV which will be
the cutoff scale on the orientifolded singularity
when the RR-tadpole of the visible branes is cancelled there \cite{Conlon:2009qa, Conlon:2010ji}.
Note that this replacement should be done just in the K\"ahler potential and
we have assumed that there are no D-branes and orientifold planes in the big-cycle $\tau_1$
and gaugino condensation arises from the $SO(8)$ pure SYM sector in $\tau_3$-cycle.

Now the $V_{U(1)}$ becomes massive, so
through $\partial_{V_{U(1)}}K = 0$ up to the gauge kinetic term and matter contribution,
one can solve the equation of motion of heavy $V_{U(1)}$, which eats $T_2$ again:
\begin{align}
V_{U(1)} + \frac{\tau_2}{\delta_{\rm GS}} &= \frac{1}{3} \frac{\alpha_2}{\delta_{\rm GS}}  
\log({\cal V}) +{\cal O}(|Q|^2).
\end{align}
One reads the SUSY-breaking order parameters from the above expression:
\begin{align}
F^{T_2} = \alpha_2 \frac{F^{T_1}}{2\tau_1} \simeq \alpha_2 m_{3/2} , \qquad
g_A^2 D_{U(1)} = -\frac{\alpha_2}{\delta_{\rm GS}} \bigg|\frac{F^{T_1}}{2\tau_1} \bigg|^2 = {\cal O}(m_{3/2}^2) .
\end{align}
Here we have used $\delta_{\rm GS} \sim \alpha_2 ={\cal O}(1/2\pi)$.
Then other moduli $T_1$ and $T_3$ are stabilized similarly to the sequestered case so long as $a \alpha_2 < 3$ \cite{Choi:2010gm}.
Otherwise, one obtains no (meta-)stable minimum, i.e. $m_{\phi_1}^2 <0$.

The matter K\"ahler metric including GM-term will be given by
\begin{align}
Z = \frac{Z_{\rm GM}}{z} = \frac{{\cal Y}(\hat{\tau}_2)}{\tau_1},~~~
\hat{\tau}_2 \equiv \tau_2 -\frac{1}{3}\beta_2 \log({\cal V}) + \delta_{\rm GS}V_{U(1)} ,
\end{align}
where ${\cal Y}(\hat{\tau}_2)$ will be a regular function in the $\tau_2 \to 0$ limit, and hence can be expanded in Taylor series:
\begin{align}
{\cal Y}(\hat{\tau}_2) \equiv \sum_{n=0} c_n \hat{\tau}_2^n , \qquad
\beta_2 = {\cal O}\bigg(\frac{1}{2\pi}\bigg).
\end{align}
In general, $\beta_2$ can be different from $\alpha_2$.
Thus, soft SUSY-breaking terms are given by
\begin{align}
m^2_i &
\simeq 
%
m_{3/2}^2 \left( 
q_i + {\cal O}(\beta_2) 
\right) , \qquad
A_{i_1 \cdots i_n}  \simeq {\cal O}(\alpha_2 -\beta_2) \times m_{3/2} , \qquad
\frac{M_{1/2}}{g_a^2} \simeq \frac{\alpha_2}{4\pi}m_{3/2} , \\
\mu &\simeq {\cal O}(\alpha_2 - \beta_2) \times zm_{3/2} , \qquad
B\mu \simeq  {\cal O}(\beta_2) \times z m_{3/2}^2  .
\end{align}
Here the first term in the scalar mass is the D-term contribution from $V_{U(1)}$: 
$\Delta_D m_i^2 = - q_i g_A^2 D_{U(1)}$, where $q_i$ is $U(1)$ charge. 

As mentioned above, in order to avoid the moduli problem, we will consider heavy soft masses 
by taking ${\cal V} = {\cal O}(10^7)$: 
\begin{align}
m_{3/2} &\sim \frac{1}{{\cal V}} \sim {\cal O}(10^{10} - 10^{11})~{\rm GeV} \gtrsim
m_{\rm soft} \sim {\cal O}(10^{9} - 10^{10})~{\rm GeV} , \\
m_{\phi_1} &\sim \frac{m_{3/2}}{\log({\cal V})^{1/2}{\cal V}^{1/2}} \sim {\cal O}(10^{6}-10^{7})~{\rm GeV}. 
\end{align}

\subsection{Dark radiation from the modulus decay}

Here we will consider the lightest modulus decay into $hh$, $ZZ$, $WW$ and $a_1 a_1$, where
$h$ is the light Higgs, $Z$ and $W$ here are the longitudinal modes
corresponding to the NG ones also in the Higgs multiplets. 
Note that the decay into SUSY particles is kinematically forbidden because no-scale structure is assumed to be
broken radiatively. 
Suppose that there is the GM-term for generating $\mu$-term and then
one finds that $q_{H_u} + q_{H_d} = 0$ is required.
In this case, as the $U(1)$ is assumed to be anomalous, 
one needs to introduce additional matter fields, 
which are vector-like under the MSSM gauge group 
and are chiral under the $U(1)$.
In addition, for all the MSSM scalar fields to be non-tachyonic at the cutoff scale,
$q_{\rm MSSM} = 0$ is required.\footnote{
This is not the case if one considers  the Froggatt-Nielsen mechanism for explaining the flavor structure of
the Yukawa coupling.
}
We will consider such a case later, introducing the QCD axion together with vector-like 
messengers charged under the $U(1)$.

Let us study the modulus decay, assuming that the energy density of the overall modulus dominates
the Universe. Similarly to the sequestered case, the modulus decays into 
a pair of axions at a rate given by (\ref{Gam2a}).  Through the GM-term, the modulus decays 
into a pair of the light Higgs bosons as well as the longitudinal modes of $Z$ and $W$ bosons at a rate given by\footnote{
In the unitary gauge, the modulus decay into longitudinal $Z$ and $W$ is induced via the kinetic mixing with the light Higgs.
}
\bea
\Gamma(\delta \phi_1 \to hh, ZZ, WW) &\simeq& \frac{z^2 \sin^2 2\beta}{48\pi}  \frac{m_{\phi_1}^3}{M_{\rm pl}^2},
\eea
where we have taken the decoupling limit, and $\tan \beta$ denotes the ratio of the up-type and down-type Higgs
boson VEVs. Note that the other heavy Higgs bosons are too heavy to be produced by the modulus decay.
One might expect that the modulus can decay into the light Higgs through $B \mu$ or $\mu$ terms,
which seems to give a larger decay rate at first sight. However, the light Higgs boson mass is due to the
cancellation between $B\mu$, $\mu$ and the scalar masses $m_{H_u,H_d}^2$, all of which are governed by 
the gravitino mass $\sim 1/{\cal V}$. Thus  the decay into $hh$ via the mass term 
is suppressed by the light Higgs mass, and is therefore sub-dominant compared to the decay via the GM-term.
%
Thus, the relevant decay modes are the decay into $hh$, longitudinal components of $ZZ$ and $WW$ through the GM-term,
and that into two axions via the kinetic term.
Summing up these contributions, one finds the total decay width of the modulus:
\bea
\Gamma_{\delta \phi_1} &\simeq & \Gamma(\delta \phi_1 \to  hh, ZZ, WW) + \Gamma(\delta \phi_1 \to 2 a_1) \\
& \simeq& \frac{z^2 \sin^2 2\beta+1}{48 \pi} \,\frac{m_{\phi_1}^3}{M_{\rm pl}^2} .
\eea
The branching fraction to the axions is given by
\begin{align}
B_a  \simeq \frac{1}{z^2 \sin^2 2\beta+1 },
\end{align}
and the reheating temperature of the modulus is estimated as
\begin{align}
\label{Td}
T_d 
&\simeq
1.9\, {\rm GeV} \times  (1-B_a)^\frac{1}{4} \left(\frac{z^2 \sin^2 2\beta+1}{4}\right)^{1/2}
\left(\frac{80}{g_*(T_d)}\right)^{1/4} 
\left(\frac{m_{\phi_1}}{10^7{\rm GeV}}\right)^{3/2}.
\end{align}
Thus, one obtains
\begin{align}
\Delta N_{\rm eff} \simeq \frac{43}{7z^2 \sin^2 2 \beta}\times \lrfp{10.75}{g_*(T_d)}{\frac{1}{3}} .
\end{align}
If we substitute $z^2=3$, $\tan \beta \approx 1$, and $g_* (T_d)=80$, we obtain $\Delta N_{\rm eff} \approx 1.0$.
Note that $\tan \beta$ should be close to $1$ in order to explain the SM-like Higgs boson 
mass, $m_h \sim 125$\,GeV~\cite{:2012gk,:2012gu}.
One can easily read the relations among $\Delta N_{\rm eff}$, $T_d$, $z$ and $m_{\phi_1}$ 
 for the non-sequestered model from  Fig. \ref{fig:dN} by noting that $z_{\rm non-seq.} \sin 2\beta \approx \sqrt{2}z_{\rm seq.}$.
Here $z_{\rm seq.}$ represents $z$ in Fig. \ref{fig:dN} and $z_{\rm non-seq.}$ is $z$ in this section.
Thus,  $z \sin 2\beta \approx 2.1 \pm 0.4$ is necessary to realize $\Delta N_{\rm eff} \approx 1.0$.
Alternatively, as before, if there are additional Higgs bosons much lighter than the modulus, 
$\Delta N_{\rm eff} \approx 1$ can be explained even for $z=1$.

Again, the decay of $\delta \phi_1 \to 2A_{\mu}$ (transverse) 
occurs through the conformal anomaly;
the partial decay width is suppressed by the 1-loop factor.

\subsection{Dark matter: Introduction of the QCD axion}

In the present case, the reheating temperature by the modulus decay is much lower than the LSP mass,
and therefore the thermal relic abundance of the LSPs is negligibly small. Also the modulus decay into the LSPs
is kinematically forbidden. Thus, the LSP abundance is too small to explain the observed dark matter.

There is another dark matter candidate, the QCD axion.
Notice that the $U(1)$ becomes an anomalous global symmetry in the low energy,
because the gauge multiplet $V_{U(1)}$ is massive in this model 
although any usual chiral multiplet does not develop the VEVs. 
Thus, the global $U(1)$ symmetry can be identified with
the Peccei-Quinn (PQ) symmetry for solving strong CP problem.

Let us include the QCD axion multiplet \cite{Murayama:1992dj, Choi:2010gm}:
\begin{align}
\Delta K &= Z_X X^{\dag} e^{2q_X V_{U(1)}}X + Z_Y Y^{\dag} e^{2q_Y V_{U(1)}}Y 
+Z_\Psi (
 \Psi^{\dag}e^{2q_\Psi V_{U(1)}} \Psi + \bar\Psi^{\dag}e^{2q_{\bar\Psi} V_{U(1)}} \bar\Psi
) 
\\
\Delta W &= \frac{X^{k+2} Y}{M_{\rm pl}^{k}} + X \Psi \bar\Psi.
\end{align}
Here $X$ becomes the QCD axion multiplet, $\Psi + \bar{\Psi}$ is charged under $U(1)_{\rm PQ}$ and 
the vector-like messenger under the MSSM gauge group.
Then effective potential is given by
\begin{align}
\nonumber
V  = m_X^2 |X_c|^2 + m_Y^2 |Y_c|^2 + \left( 
A_{XY} \frac{X_c^{k+2}Y_c}{M_*^{k}} + c.c.
\right) ~~~
\\ 
+
\frac{1}{M_*^{2k}}\left(
|X_c|^{2k+4} + (k+2)^2 |X_c|^{2k+2}|Y_c|^2
\right).
\end{align}
Here $X_c = \sqrt{Z_X}X$ and $Y_c=\sqrt{Z_Y}Y$ 
are canonically normalized fields and
$M_* = \sqrt{Z_X}M_{\rm pl}\sim M_{\rm pl}/{\cal V}^{1/3}$, assuming $Z_X = Z_Y$.
We have neglected $\tau_3$ dependence in the couplings in the scalar potential
for simplicity.
For $m_X^2 \sim q_X (\alpha_2/\delta_{\rm GS}) m_{3/2}^2 < 0$, 
the $U(1)_{\rm PQ}$ symmetry breaking is driven by tachyonic $X$.
Note that gauge invariance requires $q_X (k+2) + q_Y =0$ 
and hence $m_Y^2$ is positive, however $Y$ also acquires the VEV because of the $A_{XY}$.
Then the VEVs are found as
\begin{align}
f_a \equiv \langle X_c\rangle &\sim \left(
|m_{X}| M_*^k
\right)^{\frac{1}{k+1}} 
\sim \frac{M_{\rm pl}}{{\cal V}^{2/3}} ={\cal O}(10^{13}) ~{\rm GeV}
\qquad ({\rm for}~ k=1)
, \\
\langle Y_c\rangle &\sim 
\frac{A_{XY}}{m_{Y}} \langle X_c\rangle < \langle X_c\rangle . 
\end{align}
Here $f_a$ is the decay constant of the QCD axion and
recall that $A_{XY}$ is suppressed by the 1-loop effect.
As a consequence of $U(1)_{\rm PQ}$ breaking by $X$ VEV, 
${\rm Arg}(X)$ becomes the QCD axion, which is coupled to gluon after integrating out messenger\footnote{
Note that one obtains $F^X/X \sim A_{XY}$ and $F^Y/Y \sim m_{3/2}^2/A_{XY}$. Hence if the additional coupling of $Y$
to messenger field $\Psi' + \bar\Psi'$, $W \supset Y\Psi' \bar\Psi'$, 
the gauge mediation contribution to the soft masses would dominate over the moduli mediation
in the visible sector: $m_0 \sim A_{i_1 \cdots i_n}\sim M_{1/2} \sim m_{3/2}$. 
}.
$|X_c|$ is so-called QCD saxion and is so heavy as the gravitino mass.
Its lifetime is very short since the saxion rapidly decays into pair of the QCD axions: 
$\Gamma \sim {m_{3/2}^3}/{(64\pi f_a^2)} \sim M_{\rm pl}/{\cal V}^{5/3} \sim {\cal O}(1)$ TeV.
Thus the saxion is harmless in the cosmology.

Since the abundance of the QCD axion is given by
\begin{align}
\Omega_a h^2 \simeq 0.7 \left( \frac{f_a}{10^{12}~{\rm GeV}} \right)^{7/6} \left(\frac{\theta}{\pi} \right)^2 ,
\end{align}
a slight tuning ($\approx 10 \% $) of misalignment angle $\theta$ gives the correct dark matter abundance.
If we set the volume larger, the modulus decays after the QCD phase transition, diluting the QCD axion abundance.
In this case, the tuning of $\theta$ is relaxed, and the value of $f_a$ can be larger~\cite{Kawasaki:1995vt}.

Lastly we note that the branching fraction of the QCD axions produced by the modulus decay is suppressed by 
$f_a/M_{\rm pl}$ or the axion mass. Hence the produced QCD axions can not become the main component of the dark radiation in this model.

\section{Discussion and Conclusions}
\label{sec:5}

So far we have assumed that the lightest modulus dominates the energy density of the Universe
before it decays. If the inflation scale is sufficiently low, the modulus abundance can be suppressed by
$(H_{\rm inf}/m_\phi)^4$~\cite{Higaki:2012ba},
and it may not dominate the energy density of the Universe. The Universe is then reheated
by the inflaton decay as usual. In this case, the approximate no-scale structure has an advantage
such that the non-thermal gravitino production from the inflaton decay is suppressed~\cite{Endo:2006xg}.
The inflaton decay into the visible sector proceeds through the mixing with the modulus. For instance,
it will decay into the SM gauge sector 
through the conformal anomaly~\cite{Endo:2007ih,Endo:2007sz}.

If both the inflation scale and the inflaton are sufficiently small compared to the
modulus mass, the coherent oscillations of the modulus may not be produced at all.  For such low-scale
inflation, the successful reheating may require the inflaton to have sizable couplings to the visible sector,
for instance, the Higgs field associated with the U(1)$_{\rm B-L}$ symmetry~\cite{Nakayama:2012dw}.

We have mainly focused on the dark radiation and dark matter in this paper, but successful cosmology
requires the generation of the baryon asymmetry as well.
Since the modulus decay produces a huge amount of entropy,  any pre-existing
baryon asymmetry is diluted. Therefore, an efficient baryogenesis is needed to account for the observed 
baryon asymmetry. In Ref.~\cite{Higaki:2012ba}, the Affleck-Dine mechanism~\cite{Affleck:1984fy}
was considered in detail, and it was shown that a right amount of
the baryon asymmetry can be indeed created.

We showed that the dark radiation is generated from the decay of
the lightest modulus in the LARGE volume scenario,
studying both cases of sequestered LVS and non-sequestered one.
In both models, 
natural and robust candidate of the dark radiation is the ultralight string theoretic axion, 
which is the superpartner of the lightest modulus.
This is because of the no-scale structure and a large VEV of the modulus.
As a result, the abundance of axions from the modulus decay via the axion kinetic term competes nicely
that of the Higgses produced by the decay via the GM-term, which is necessary for generating higgsino mass.
According to LVS, the existence of the dark radiation is quite plausible outcome.

Depending on the model in string theories, additional ultralight fields may appear.
For instance, with respect to the moduli stabilization or an uplifting potential via gaugino condensations, 
a hidden sector with matter
would be also available (by adding the closed/gauge fluxes to the bulk/brane even if the cycle is not rigid one).
For such a case, ultralight NG fields might appear
by the breakdown of (approximate) chiral global symmetries which are an $U(1)$ or $SU(N_{\rm flavor})$.
This will be also the case for the uplifting potential originating from a dynamical SUSY-breaking sector where
such global symmetries are broken down without any generation of gaugino condensations, like in the ISS model
\cite{Intriligator:2006dd}.
In these cases, such massless modes would be also candidates of dark radiation from the lightest modulus decay
since the decay fraction of the lightest modulus into the NG modes can be comparable to the main decay mode 
in Eq. (\ref{maindecay}) \cite{Cicoli:2010ha}.
Perhaps, their superpartners could become a component of the dark matter candidate
since they would also acquire TeV scale SUSY-breaking soft masses.

Further model-dependent study of dark radiation will be significant for 
viable cosmology derived from string theories.

\vspace{5mm}

{\bf Note added:}

While discussing on the dark radiation in the LVS, we were told from Joe Conlon that he and his collaborators 
were working on the closely related topics \cite{Cicoli:2012aq}.

\section*{Acknowledgments}
Authors would like to thank Joe Conlon for fruitful discussions and useful 
comments on the soft scalar mass in the large volume scenario, and
Kiwoon Choi for his comments on the radiative breaking of the sequestering. 
F.T. thanks Kwang Sik Jeong for discussion.
Authors would like to thank also 
the 3rd UTQuest workshop ExDiP 2012 Superstring Cosmophysics held at
Obihiro in Japan, where the present work was initiated. 
This work was supported by the Grant-in-Aid for Scientific Research
on Innovative Areas (No.24111702, No.21111006 and No.23104008) [FT],
Scientific Research (A) (No.22244030 and No.21244033 [FT]), and JSPS
Grant-in-Aid for Young Scientists (B) (No.24740135) [FT].  This work
was also supported by World Premier International Center Initiative
(WPI Program), MEXT, Japan.

\appendix

\section{Derivation of $\Delta N_{\rm eff}$}
\label{dNeff}

In this Appendix we express $\Delta N_{\rm eff}$ in terms of the 
branching fraction of the axion production $(B_a)$ and the relativistic degrees of freedom 
at the modulus decay $(g_*(T_d))$.

The effective number of neutrinos $N_{\rm eff}$ is equal to $3$ in the standard
cosmology\footnote{Here we do not take account of the electron positron annihilation. This is
valid when one estimates the $^4$He abundance, while it should be taken into account
(although the difference is well below the current sensitivity) when one discusses the effect
on CMB.}. If there are additional relativistic degrees freedom, it is customary to quantify its energy density
in the unit of one neutrino species (e.g., $\nu_e$ + $\bar{\nu}_e$) when the neutrinos decouple
from plasma, i.e., at $T \sim$ a few MeV. That is, $\Delta N_{eff}$ is defined as follows.
\bea
\Delta N_{\rm eff}  & \equiv &\left.\frac{\rho_{DR}}{\rho_{\nu1}}\right|_{\nu \,\,decouple},
\eea
where $\rho_{\nu1}$ represents the energy density of one neutrino species,
and $\rho_{DR}$ denotes the dark radiation energy density.

Let us assume that the modulus dominates the energy density of the Universe and 
decays into massless axions and the SM particles with the reheating temperature $T_d$. 
Let $g_*(T_d)$ be the effective light degrees of freedom at $T=T_d$.
The energy density of non-thermalized axions and thermalized radiation evolves as
\bea
\rho_a(t) &\propto& a^{-4}(t) \\
\rho_{SM}(t) &\propto& g_*^{-1/3}(t) a^{-4}(t) 
\eea
where $a(t)$ denotes the scale factor. Note that it is the entropy in the comoving volume,
$\sim a(t)^3 g_*(T) T^3$, that is conserved when the number of light degrees of freedom changes.
Thus, the ratio changes with time as
\beq
\left.\frac{\rho_a}{\rho_{SM}}\right|_{T=T_1} = \lrfp{g_*(T_1)}{g_*(T_2)}{\frac{1}{3}} 
\left.\frac{\rho_a}{\rho_{SM}}\right|_{T=T_2}  
\eeq

We define $B_a$ as the branching fraction of the modulus into a pair of axions.
Then the energy densities of the axion and the SM radiation at the decay is
\beq
\left.\frac{\rho_a}{\rho_{SM}}\right|_{T=T_d} \;\simeq\; \frac{B_a}{1-B_a}
\eeq
as long as the pair production is the dominant production process of the axions. 

Let us now derive the expression for $\Delta N_{eff}$ as follows.
\bea
\Delta N_{\rm eff}  & \equiv &\left.\frac{\rho_a}{\rho_{\nu1}}\right|_{\nu \,\,decouple}
= \left.\frac{\rho_{SM} }{\rho_{\nu1}}\right|_{\nu \,\,decouple}  \left.\frac{\rho_{a} }{\rho_{SM}}\right|_{\nu \,\,decouple}\\
&=& \frac{43}{7} \times \lrfp{10.75}{g_*(T_d)}{\frac{1}{3}} \left.\frac{\rho_a}{\rho_{SM}}\right|_{T=T_d}  \\
&=&  \frac{43}{7} \times \lrfp{10.75}{g_*(T_d)}{\frac{1}{3}} \frac{B_a}{1-B_a}.
\eea
For instance, if  we substitute $B_a = 0.25$ and $g_*=80$, we obtain $\Delta N_{\rm eff} \approx 1.05$.




\end{document}